# Phase diagram and upper critical field of homogenously disordered epitaxial 3-dimensional NbN films

Mintu Mondal[a*], Madhavi Chand[a], Anand Kamlapure[a], John Jesudasan[a], Vivas C. Bagwe[a], Sanjeev Kumar[a], Garima Saraswat[a], Vikram Tripathi[c] and Pratap Raychaudhuri[a†]

[a] *Department of Condensed Matter Physics and Materials Science, Tata Institute of Fundamental Research, Homi Bhabha Road, Mumbai 400005, India*

[c] *Department of Theoretical Physics, Tata Institute of Fundamental Research, Homi Bhabha Road, Mumbai 400005, India*

We report the evolution of superconducting properties with disorder, in 3 dimensional homogeneously disordered epitaxial NbN thin films. The effective disorder in NbN is controlled from moderately clean limit down to Anderson metal-insulator transition by changing the deposition conditions. We propose a phase diagram for NbN in temperature-disorder plane. With increasing disorder we observe that as $k_F l \rightarrow 1$ the superconducting transition temperature ($T_c$) and normal state conductivity in the limit T→0 ($\sigma_0$) go to zero. The phase diagram shows that in homogeneously disordered 3-D NbN films, the metal-insulator transition and the superconductor-insulator transition occur at a single quantum critical point, $k_F l \sim 1$.

* mondal@tifr.res.in

† pratap@tifr.res.in

## 1. INTRODUCTION:

The effect of disorder on superconducting properties of a material is one of the most challenging problems in condensed matter physics. In 1959 Anderson [1] proposed that the disorder arising from non magnetic impurities does not affect the superconducting properties. It was however understood later that Anderson theorem is only applicable for weakly disordered systems [2]. In case of strongly disordered systems, as the disorder strength increases, the Bloch states become localized and give rise to Anderson Metal-Insulator transition (MIT). Close to the MIT the superconductivity is controlled by the competition between the superconducting energy gap and the localization length ($l_{loc}$). If the superconducting energy gap ($\Delta$) is larger than the energy level spacing ($\delta\varepsilon$) of the confined eigenstates within a length $l_{loc}$, the pairing of localized eigenstates is favorable and leads to local superconductivity on the scale of the $l_{loc}$. Eventually through Josephson coupling between these localized superconducting regions, it gives a phase coherent superconducting ground state even in an insulating system. In this scenario, superconductivity persists in the insulating regime [3,4], as long as $N(0)l_{loc}^3\Delta > 1$, where *N(0)* is the density of states at Fermi level. Eventually, as the disorder is further increased the system is expected to undergo transition from a superconducting to insulating ground state. However there are some experimental evidences [5,6] that in many homogeneously disordered system the superconductivity is completely suppressed in the metallic side of the metal-insulator transition while the system is still a metal. It has been argued that [7] the suppression of $T_c$ in the disordered metallic state can arise from the increase in electron-electron interactions due to loss of effective screening in the presence of disorder. The enhanced coulomb interaction competes with the electron-electron (e-e) attractive interaction in the Cooper channel leading to the destruction of superconductivity in the metallic state.

Here we study the evolution of superconducting transition temperature ($T_c$) and upper critical field ($H_{c2}$) as a function of disorder in homogeneously disordered epitaxial 3D-NbN[8]. All the films are in the 3-D limit with thickness (>50nm) much larger than the dirty limit coherence length ($\xi$~5nm). The disorder spans a large range from moderately clean ($k_Fl$~10) to the Mott limit of minimum metallic conductivity, i.e. $k_Fl$~1. In this system, we present evidence of a coinciding Metal-Insulator and Superconductor-Insulator transition as the disorder approaches $k_Fl$~*1*. In this range of disorder, we observe that the Ginzburg-Landau coherence length ($\xi_{GL}$) estimated from upper critical field ($H_{c2}(0)$) increases by a factor of 2 with increasing disorder in contrary to the expectation in the weak disorder regime where $\xi_{GL}$ decrease with increasing disorder.

## 2. EXPERIMENTAL DETAILS:

Epitaxial NbN films were grown on (100) oriented single crystalline MgO substrates using reactive d.c. magnetron sputtering of Nb in $Ar/N_2$ gas mixture. The effective disorder of these films was controlled by changing the sputtering power or the $Ar/N_2$ ratio, which allowed us to grow films over a large range of disorder. Details of preparation and characterization of these samples have been reported in reference [8 ]. The temperature dependence of resistivity (ρ) was measured using standard four probe technique. To determine $H_{c2}(0)$, we have measured resistivity in presence of magnetic field. The $H_{c2}(T)$ was determined from $\rho(T,H)$ versus $T$ data taken at different applied magnetic field. Hall coefficient ($R_H$) was measured using a standard four-probe ac technique on films patterned in Hall-bar geometry. $R_H$ was calculated from Hall voltage deduced from reversed field sweeps from +12 to −12 T after subtracting the resistive contribution. The thickness of the films was measured using a stylus profilometer. For each film $k_Fl$ was determined from the ρ(285K) and $R_H$(285K) using the free electron formula $k_F l = \left\{(3\pi^2)^{2/3}\hbar[R_H(285K)]^{1/3}\right\}/\left[\rho(285K)e^{5/3}\right]$, where $e$ is the electron charge.

## 3. RESULTS AND DISCUSSION:

Figure 1(a) and 1(b) show the resistivity as a function of temperature and figure 1(c) shows the conductivity ($\sigma$) as a function of temperature for a set of films with $k_F l$ ranging from $k_F l$ =1.24 to $k_F l=10.12$. The superconducting transition temperature, $T_c$, of the films is determined from the temperature at which resistance falls to 10% of its normal state value. All the films with $k_F l \leq 8.13$ show a positive $d\sigma/dT$ up to room temperature. Here we define $\sigma_0$ as the d.c. conductivity just above $T_c$. For films $k_F l \leq 8.13$, $\sigma_0$ is the minimum conductivity. All our films show finite $\sigma_0$ including the highest disordered film with $k_F l=1.24$, indicates that all the films are in metallic regime.

Figures 2(a) and 2(b) show the variation of $T_c$ and $\sigma_0$ as function of $k_F l$ for all the films. From the trend of variation of $T_c$ and $\sigma_0$ with $k_F l$, it is clear that both $T_c$ and $\sigma_{0\,are}$ going to zero as $k_F l \rightarrow 1$. Thus the MIT and the SIT in this system coincides at a single quantum critical point, $k_F l \sim 1$. Based on $T_c$ vs $k_F l$ and $\sigma_0$ vs $k_F l$ data, we propose the phenomenological phase diagram in Figure 3.

The nearly exact coincidence of MIT and SIT observed in NbN is intriguing. While superconductivity can get destroyed either in the insulating or the metallic side of the MIT depending on the relative contribution of localization and *e-e* interactions, so far no theory explicitly predict that these two transitions will coincide at a single point. Previous studies have indicated that both localization and Coulomb effect are present in our system [8]. While it is possible in principle for this coincidence to be purely accidental, such a possibility seems extremely unlikely. On the other hand a coinciding transition has also been observed before in epitaxial boron-doped diamond films [9]. All previous studies on strongly disordered superconductors have been performed on amorphous or granular films. It is therefore possible that homogeneously disordered epitaxial superconductors behave qualitatively in a different way compared to their granular/amorphous counterparts.

Figure 4 (a) shows the $\rho(H,T)$ as function of $T$ for a film with $T_c$~4.2K at different magnetic field. $H_{c2}(T)$ is determined the ρ(H,T) vs T data where ρ(H,T) goes to 50% of its normal state resistivity just above $T_c$. $H_{c2}(T)$ as function of $T$ for a series of films is shown in figure 4(b). Since all our films are in the dirty limit, $l<<\xi_{GL}$, we have estimated $H_{c2}(0)$ and $\xi_{GL}$ from the dirty limit relation, [10]

$$H_{c2}(0) = 0.69 T_c \left.\frac{dH_{c2}}{dT}\right|_{T=T_c} \quad \text{and} \quad \xi_{GL} = \left[\frac{\phi_0}{2\pi H_{c2}(0)}\right]^{1/2} \text{-----(1)}$$

Figure 4(c) and (d) show the variation of $H_{c2}(0)$ and $\xi_{GL}$ as a function of $T_c$ obtained using equation (1). It is interesting to note that $\xi_{GL}$ increases monotonically by a factor of 2 while $T_c$ decreases by one order of magnitude. In case of weak disorder regime where Anderson theorem is valid, $\xi_{Gl}$ decreases with increasing disorder due to the reduction in electronic mean free path ($l$). In our case however, the BCS coherence length ($\xi_{BCS}$) itself increases due to the decrease in $T_c$. Therefore $\xi_{Gl} = (\xi_{BCS} l)^{1/2}$, is determined by a competition between $\xi_{BCS}$ and $l$. We have independently estimated $\xi_{BCS}$ from the measured superconducting energy gap($\Delta_0$) from tunneling measurement, Fermi velocity($v_F$) and electronic mean free path ($l$) using following dirty limit BCS relation,

$$\xi_{BCS} = \frac{\hbar v_F}{\pi \Delta(0)} \quad \text{and} \quad \xi_{GL} = (l \xi_{BCS})^{1/2} \text{---------(2)}$$

The calculated numbers reproduce the experimental trend while the estimated coherence length ($\xi_{GL}$) using dirty limit BCS relation is approximately twice the experimentally measured value for the respective disordered film. The relative insensitivity of $\xi_{Gl}$ to disorder is because the increase in $\xi_{BCS}$ due to reducing $T_c$ is partially compensated by the decrease in mean free path.

Further investigations on the superconducting properties show that for low disorder NbN follows the conventional BCS behavior [11,12]. However, as we approach the SIT the superconducting state is characterized by a gradual loss in the coherence peak in the tunneling density of states and a linear variation of the superfluid density with temperature over a large temperature range. The unusual behavior in tunneling density of states and superfluid density will be discussed in a future publication [13].

## 4. CONCLUSION:

We have established a phase diagram for homogeneously disordered 3 dimensional NbN films where MIT and SIT coincide at single quantum critical point, $k_F l \sim 1$. At this moment it is not clear whether this coincidence is merely accidental or has a deeper significance. We have seen that the coherence length ($\xi_{GL}$) monotonically increases with disorder in contrast with expected behavior from weak disorder limit.

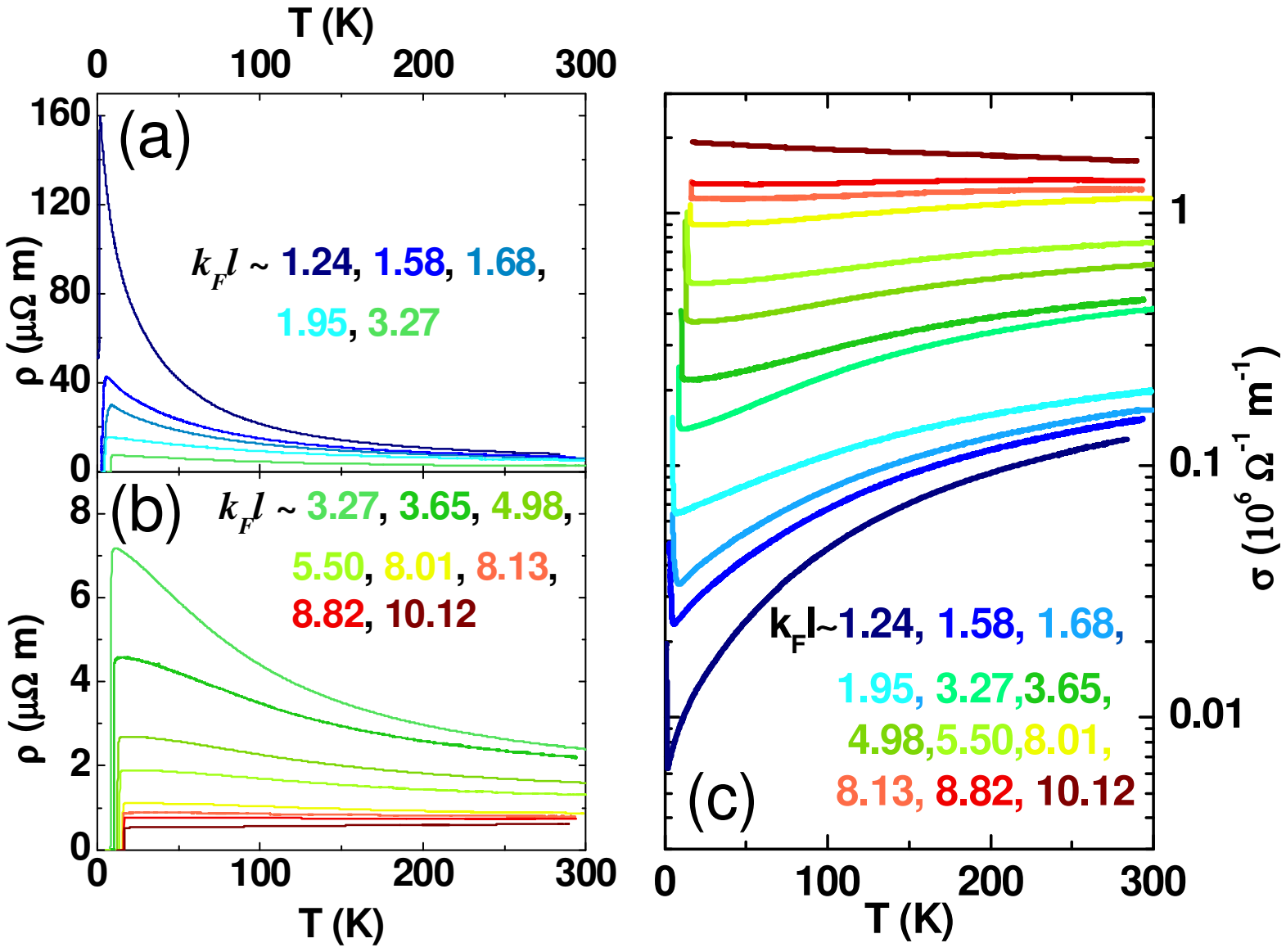

***Figure 1.** (a) Resistivity (ρ) as a function temperature (T) for a set of films with $k_F l$ value, from $k_F l$ =1.24 to $k_F l$ =3.27. (b) Resistivity (ρ) as a function temperature (T) for a set of films with $k_F l$ value 3.27 to $k_F l$ =10.12. Note that the resistivity data for the films with $k_F l$=3.27 is plotted in both figure 1(a) and 1(b). (c) Conductivity (σ) as a function temperature (T) for a set of films with $k_F l$ values from $k_F l$=1.24 to $k_F l$=10.12.*

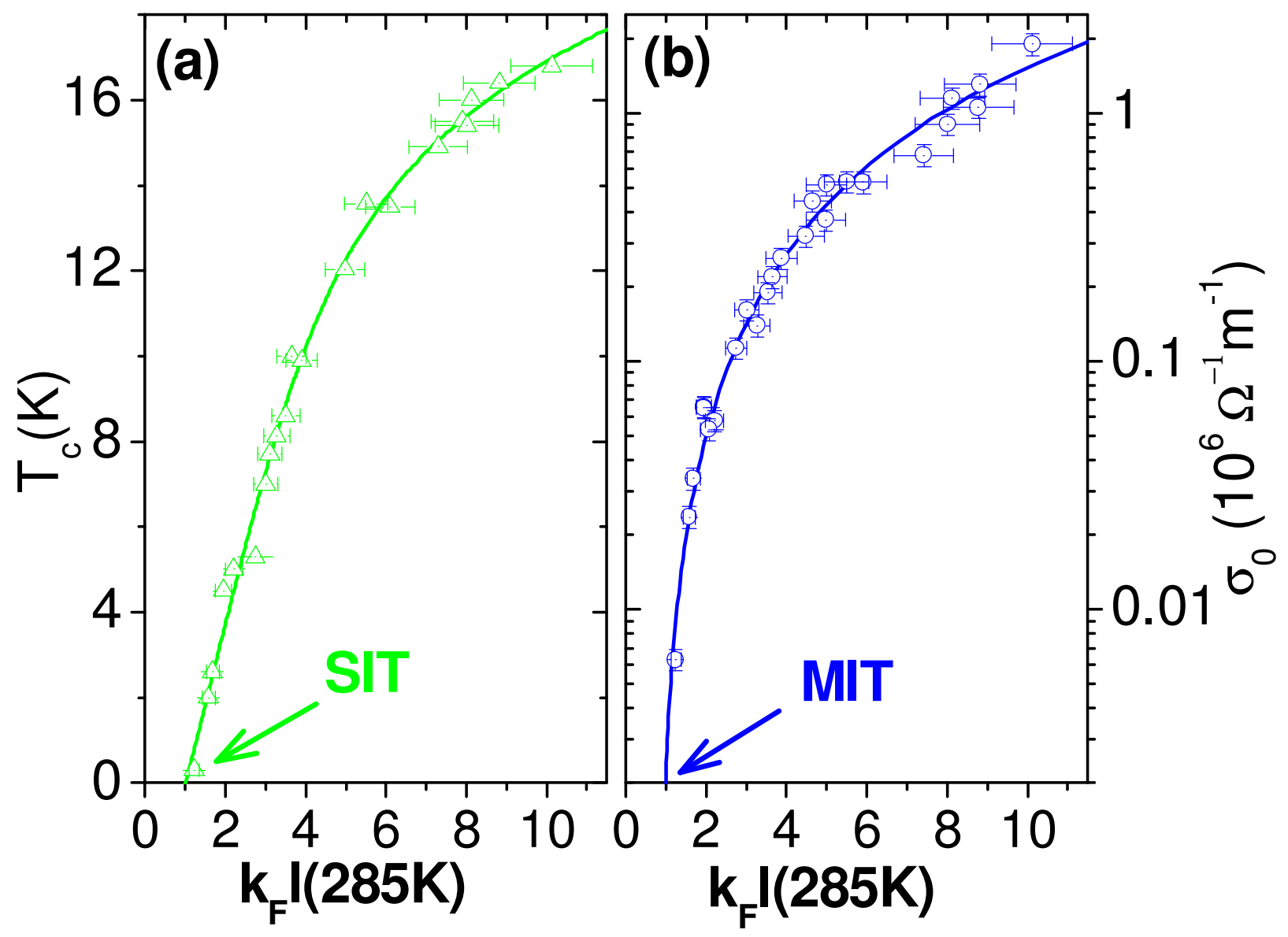

***Figure 2.** (a) $T_c$ vs $k_Fl$ for with $k_Fl$~1.23 to 10.12, showing $T_c \rightarrow 0$ as $k_Fl \rightarrow 1$; (b) $\sigma_0$ vs $k_Fl$ where $\sigma_0$ is in log scale showing $\sigma_0 \rightarrow 0$ as $k_Fl \rightarrow 1$.*

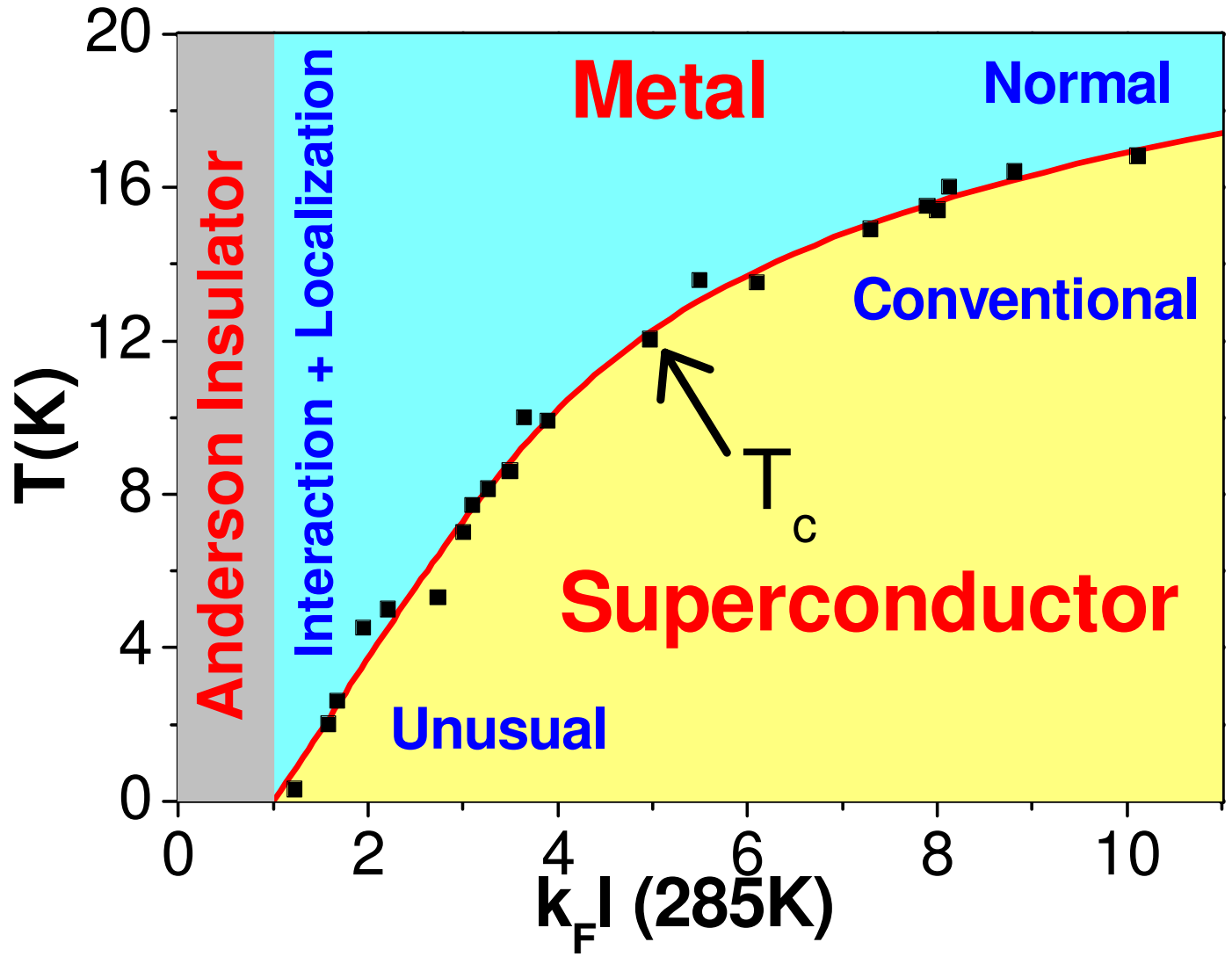

**Figure 3** ***Phase diagram of homogenously disordered epitaxial 3 dimensional NbN films. The Metal-Insulator transition (MIT) and Superconductor-Insulator transitions (SIT) coincide at a single quantum critical point, $k_F l$ ~ 1.***

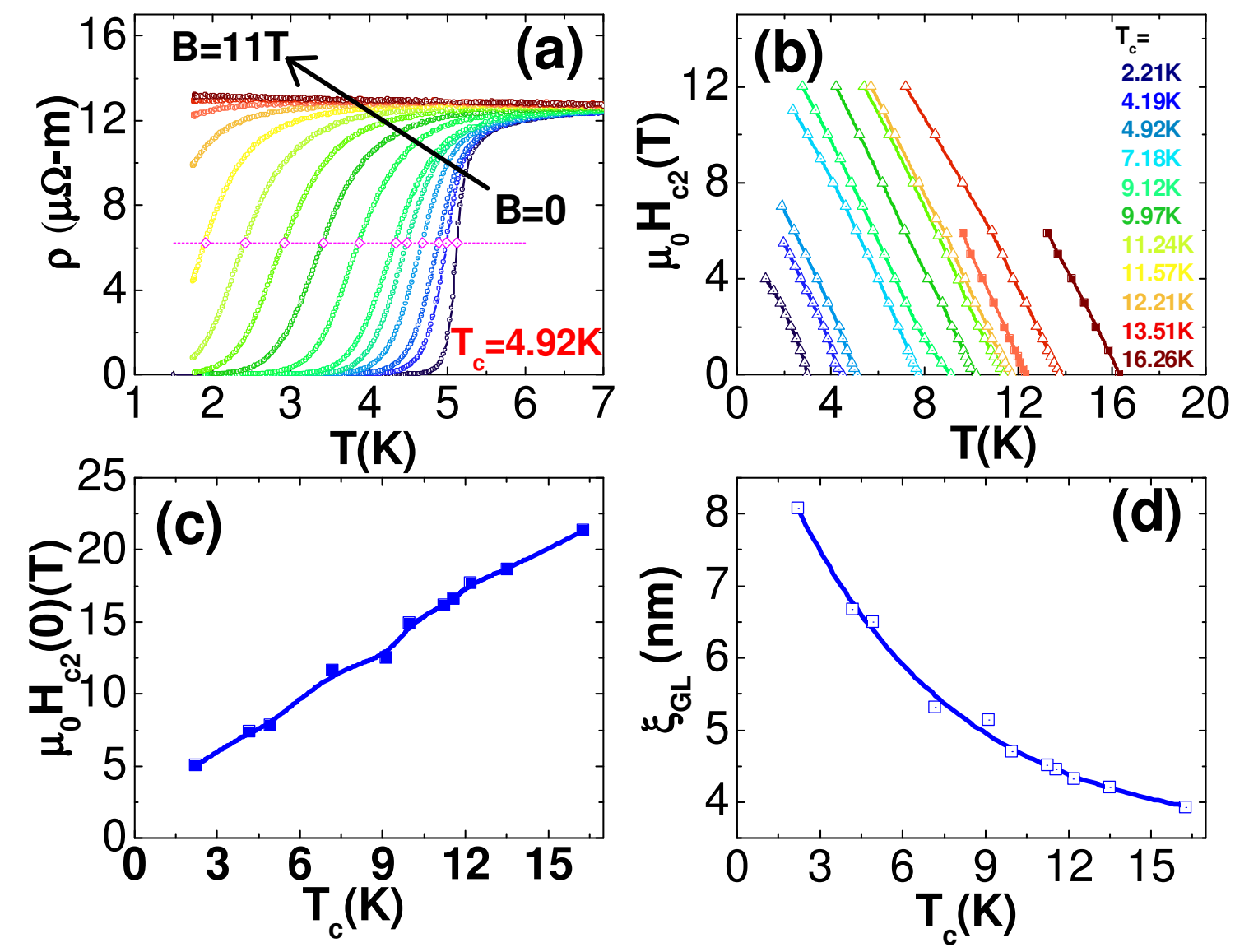

***Figure 4. (a) Resistivity (ρ) as a function temperature (T) for film of $T_c$~4.92K in different magnetic fields. The successive magnetic field values are 0, 0.25, 0.5, 1, 1.5, 2, 2, 3, 4, 5, 6, 7, 8, 9, 10 and 11T; (b) The upper critical field ($H_{c2}$) as a function of temperature (T) for series of films with different superconducting transition temperature ($T_c$).*** **(b) *$H_{c2}(0)$* as a function of *$T_c$* ; (c) G-L Coherence length *($\xi_{GL}$)* as a function of *superconducting transition temperature ($T_c$)*.**

## References:

[1] P.W. Anderson, J. Phys. Chem. Solids 11 (1959) 26.

[2] M. Ma and P. A. Lee, **Phys. Rev. B 32, 5658 (1985) .**

[3] B. Sacépé, C. Chapelier, T. I. Baturina, V. M. Vinokur, M. R. Baklanov, and M. Sanquer, Phys. Rev. Lett. **101**, 157006 (2008); G. Sambandamurthy, L. W. Engel, A. Johansson, E. Peled, and D. Shahar, Phys. Rev. Lett. **94**, 017003 (2005).

[4] M. V. Feigel'man, L. B. Ioffe, V. E. Kravtsov, and E. A. Yuzbashyan, Phys. Rev. Lett. **98**, 027001 (2007); A. Ghosal, M. Randeria, and N. Trivedi, Phys. Rev. Lett. **81**, 3940 (1998); Phys. Rev. B **65**, 014501 (2001).

[5] T. Furubayashi, N. Nishida, M. Yamaguchi, K. Morigaki and H. Ishimoto, Solid State Common. 55 (1985) 513

[6] B.J. Bishop, E.G. Spencer and R.C. Dynes, Solid State Electron. 28 (1985) 73; G. Hertel, D.J. Bishop, E.G. Spencer, J.M. Rowell and R.C. Dynes, Phys. Rev. Lett.50 (1983) 743.

[7] A. Finkelshtein, JETP Lett. **45**, 46 (1987).

[8] S. P. Chockalingam, M. Chand, J. Jesudasan, V. Tripathi, and P. Raychaudhuri,Phys. Rev. B 77, 214503(2008); M. Chand, A. Mishra, Y. M. Xiong, A. Kamlapure, S. P. Chockalingam, J.Jesudasan, V. Bagwe, M. Mondal, P. W. Adams, V. Tripathi, and P. Raychaudhuri,Phys. Rev. B **80**, 134514 (2009).

[9] T. Klein, P. Achatz, J. Kacmarcik, C. Marcenat, F. Gustafsson, J. Marcus, E. Bustarret, J. Pernot, F. Omnes, Bo E. Sernelius, C. Persson, A. Ferreira da Silva, and C. Cytermann. Phys. Rev. B **75**, 165313(2007)

[10] N. R. Werthamer, E. Helfland, and P. C. Honenberg, Phys. Rev. **147**, 295 (1966)

[11] Anand Kamlapure, Mintu Mondal, Madhavi Chand, Archana Mishra, John Jesudasan, Vivas Bagwe, L. Benfatto, Vikram Tripathi, Pratap Raychaudhuri, APL 96, 072509 (2010)

[12] S. P. Chockalingam, M. Chand, A. Kamlapure, J. Jesudasan, A. Mishra, V. Tripathi, and P. Raychaudhuri, Phys. Rev. B **79**,094509 (2009).

[13] Mintu Mondal et al, **arXiv:1006.4143**